\documentclass[%
 aip,
 amsmath,amssymb,
 print,%
]{revtex4-1}

\usepackage{graphicx}
\usepackage{dcolumn}
\usepackage{bm}

\begin{document}


\title{Two electrostatic gyrokinetic models derived by two different perturbative methods}

\author{Shuangxi Zhang}
\email{zshuangxi@gmail.com}
\affiliation{Graduate School of Energy Science, Kyoto University, Uji, Kyoto 611-0011, Japan.}


\date{\today}

\begin{abstract}
This paper presents two different electrostatic gyrokinetic models derived through two different perturbative methods. One of the two models is just the conventional electrostatic gyrokinetic model, the derivation of which is repeated using the Lie transform perturbative method. One term is rectified in the derivation of the conventional model. To derive the other model, we use a new method, which is based on the covariant transform formula of the differential 1-form. The new method doesn't split the coordinate transform into the guiding-center transform and the gyrocenter transform. It carries out the coordinate transform up to the order equaling that of the amplitude of the perturbative wave. Compared with the conventional model, the finite Larmor radius terms are completely removed from the orbit equations of the new one, making it simpler for the numerical application.
\end{abstract}

\maketitle

\section{Introduction}\label{sec1}

The calculation of Vlasov equation in the full-orbit kinetic numerical simulation of magnetized plasmas costs most of the simulation time to solve the evolution of a six-dimensional distribution function\cite{2010garbet1, 2012krommes, 1983wwlee,1983dubin, 1988hahm, 2008idomura, 2000jenko, wwleejcp1987, 1993parker, 2009peeters, 1993hammett}. To reduce the simulation time, the full-orbit kinetic model is simplified by assuming that the distribution function on gyrocenter coordinate is gyrocenter symmetric. Alternatively, the distribution function on gyrocetner coordinate is independent of the gyrophase. So the dimension of the phase space in Vlasov equation is reduced from six to five by transforming the particle coordinate to the gyrocenter coordinate with the gyrophase reduced due to the gyrocenter symmetry of the distribution. This coordinate transform is determined by the fact that the magnetic moment is an adiabatic quantity in terms of low-frequency perturbations\cite{1982frieman, 1990brizard, 2007brizard1, 2000qinhong1, 2010scott, 2000sugama, 2006shaojiewang, 2016tronko}, and results further in  that the orbit equations on the gyrocenter coordinate are independent of the gyrophase.

Cary-Littlejohn single-parameter Lie transform perturbative method is usually applied to obtaining the coordinate transform\cite{1983dubin,1988hahm,1990brizard}. If the system includes multiple parameters representing the multiple characteristic scales contained by the perturbations, this single-parameter method can be repeated for multiple times, with one time to handle one parameter. The principle of the Lie transform method used in the gyrokinetic theory is to get a new Lagrangian 1-form independent of the gyrophase by introducing appropriate generators into the formal formula of the new Lagrangian 1-form, which is the solution of the covariant transform formula of the differential 1-form. These generators in turn determine the coordinate transform.
Since the perturbation in magnetized plasmas includes several scales, e.g., the Larmor radius, the amplitude and the spatial scale of the perturbations, the Lie transform method used for the conventional electrostatic gyrokinetic model (CEGM) contains two parameters, and the formal formula of the new Lagrangian 1-form is adopted as $\Gamma  = \exp \left( { -\varepsilon_2 {L_{{{\bf{g}}_2}}}} \right)\exp \left( -\varepsilon_1 {{L_{{{\bf{g}}_1}}}} \right)\gamma \left( {\bf{Z}} \right)$, with $\varepsilon_1, \mathbf{g}_1$ for the guiding-center transform and $\varepsilon_2, \mathbf{g}_2$ for the gyrocenter transform. $O(\varepsilon_1)$ is the order of the magnitude of the normalized Larmor radius and $O(\varepsilon_2)$ is the order of the normalized amplitude of the perturbative wave.

In this paper, we present a new method, which derives the coordinate transform by using the covariant transform formula of the differential 1-form as a covariant vector, rather than the formal formula of the new Lagrangian 1-form. Since the covariant transform formula associates the formulas of the new 1-form, the old 1-form and the coordinate transform together, it is possible to derive a new 1-form independent of the gyrophase based on this formula by straightforwardly choosing appropriate coordinate transform. It's found that it doesn't need to split the coordinate transform into two independent ones for each parameter. The result shows that the Finite Larmor Radius terms can be completely removed from the orbit equations by the new methods. Therefore, it's simpler for the numerical application, compared with CEGM.

The rest of the paper is arranged as follows. In Sec.(\ref{sec2}), the origional Lagrangian differential one-form, which determines the orbit equations on particle coordinates, is introduced and normalized. And the multiple scales are explained. The detailed derivation of CEGM is given in Sec.(\ref{sec3}) in terms of the low-frequency electrostatic perturbations. Sec.(\ref{sec5}) presents the derivation of the new electrostatic gyrokinetic model by using the new method based on the covariant transform formula.
Sec.(\ref{sec6}) is the summary and discussion.

\section{The introduction of the Lagrangian differential 1-form and the multiple scales of the perturbations}\label{sec2}

The Lagrangian differential 1-form which determines the orbit of a test charged particle in magnetized plasmas is
\begin{equation}\label{a119}
\gamma = \left( {q{\bf{A}}\left( {{{\bf{x}}}} \right) + m{\bf{v}}} \right)\cdot d{\bf{x}} - (\frac{1}{2}m{v^2}+q \phi(\mathbf{x},t))dt.
\end{equation}
$(\mathbf{x},\mathbf{v})$ is the full particle coordinate frame.
The test particle is chosen from a thermal equilibrium plasma ensemble, e.g., the thermal equilibrium plasma in tokamak. Therefore, $\mathbf{A},\mathbf{v},\mathbf{x},t,\mathbf{B}, \phi$ can be normalized by $A_0\equiv B_0 L_0,v_t,L_0,L_0/v_t,B_0,A_0 v_t$, respectively. $B_0, L_0$ are the characteristic amplitude and spatial length of the magnetic field, respectively. $v_t$ is the thermal velocity of the particle ensemble which contains the test particle.

The detailed normalization procedure of $\gamma$ is given as follows. First, both sides of Eq.(\ref{a119}) are divided by $m{v_t}{L_0}$. The first term of RHS of Eq.(\ref{a119}) is $\frac{{q{A_0}}}{{m{v_t}}}\frac{{{\bf{A}}\left( {{{\bf{x}}}} \right)}}{{{A_0}}}\cdot\frac{{d{{\bf{x}}}}}{{{L_0}}}$, which is further written as $\frac{1}{\varepsilon }{\bf{A}}\left( {{{\bf{x}}}} \right)\cdot d{{\bf{x}}}$, with the replacement: $\varepsilon  \equiv \frac{{m{v_t}}}{{q{A_0}}}$, $\frac{{{\bf{A}}\left( {{{\bf{x}}}} \right)}}{{{A_0}}} \to {\bf{A}}\left( {{{\bf{x}}}} \right),\frac{{d{{\bf{x}}}}}{{{L_0}}} \to d{{\bf{x}}}$. Other terms can be normalized in the same way. Eventually, we could derive a normalized Lagrangian 1-form like
\begin{equation}
\frac{{{\gamma}}}{{m{v_t}{L_0}}} = \left( {\frac{1}{\varepsilon }{\bf{A}}\left( {\bf{x}} \right) + {\bf{v}}} \right)\cdot d{\bf{x}} - (\frac{1}{2}{{\bf{v}}^2} + \frac{1}{\varepsilon }\phi \left( {{\bf{x}},t} \right))dt,
\end{equation}
Now, multiplying both sides by $\varepsilon$, and rewriting $\frac{{{\varepsilon \gamma}}}{{m{v_t}{L_0}}}$ to be $\gamma$, the normalized 1-form is given as follows
\begin{eqnarray}\label{g2}
\gamma  = \left( {{\bf{A}}\left( {\bf{x}} \right) + \varepsilon {\bf{v}}} \right)\cdot d{\bf{x}} - \left( {\varepsilon \frac{{{v^2}}}{2} + \phi \left( {{\bf{x}},t} \right)} \right)dt.
\end{eqnarray}
Since a constant factor $\frac{\varepsilon }{{m{v_t}{L_0}}}$ doesn't change the dynamics determined by the Lagrangian 1-form, the Lagrangian 1-form given by Eq.(\ref{g2}) possesses the same dynamics with that given by Eq.(\ref{a119}).

The velocity can be written in cylindrical coordinates, by transforming $(\mathbf{x},\mathbf{v})$ to $(\mathbf{x},u_1,\mu_1,\theta_1)$, where $u_1$ is parallel velocity and $\mu_1$ is magnetic moment, with their definitions being $u_1\equiv \mathbf{v}\cdot \mathbf{b}$ and $\mu_1\equiv v_\perp^2/2B(\mathbf{x})$. The unit vector of the perpendicular velocity is ${\widehat {\bf{v}}_ \bot } \equiv \left( {{{\bf{e}}_1}\sin \theta  + {{\bf{e}}_2}\cos \theta } \right)$. $(\mathbf{e}_1,\mathbf{e}_2,\mathbf{b})$ are orthogonal mutually and $\mathbf{b}$ is the unit vector of the equilibrium magnetic field. After this transformation, $\gamma$ becomes
\begin{eqnarray}\label{a11}
\gamma  &=& \left( {{\bf{A}}\left( {\bf{x}} \right) + \varepsilon {u_1}{\bf{b}} + \varepsilon \sqrt {2B({\bf{x}}){\mu _1}} {{\widehat {\bf{v}}}_ \bot }} \right)\cdot d{\bf{x}} \nonumber \\
&&- \left[ {\varepsilon \left( {\frac{{u_1^2}}{2} + {\mu _1}B({\bf{x}})} \right) + \phi \left( {{\bf{x}},t} \right)} \right]dt
\end{eqnarray}
which can be splitted into three parts as
\begin{equation}\label{a12}
{\gamma _0} = {{\bf{A}}}\left( {\bf{x}} \right)\cdot d{\bf{x}},
\end{equation}
\begin{equation}\label{a13}
{\gamma _1} = \varepsilon\left( {{u_1}{\bf{b}} + \sqrt {2B({\bf{x}}){\mu _1}} {{\widehat {\bf{v}}}_ \bot }} \right)\cdot d{\bf{x}} - \varepsilon\left( {\frac{{u_1^2}}{2} + {\mu _1}B({\bf{x}})} \right)dt,
\end{equation}
\begin{equation}\label{a14}
{\gamma _{\sigma } } =  - \phi \left( {{\bf{x}},t} \right)dt.
\end{equation}
The $\mathbf{X}$ components in $\gamma_1$ can be decomposed into  the parallel and perpendicular parts as  $\gamma_{1\mathbf{x}\parallel}=\varepsilon u\mathbf{b}$ and $\gamma_{1\mathbf{x}\perp}=\varepsilon \sqrt {2B({\bf{x}}){\mu _1}} {{\widehat {\bf{v}}}_ \bot }$.

$\theta$ is a fast variable and the term depending on $\theta$ in Eq.(\ref{a13}) is ${\varepsilon }\sqrt {2\mu_1 B({\bf{x}})} {\widehat {\bf{v}}_ \bot }\cdot d{\bf{x}}$ possessing the order $O(\varepsilon)$. $\theta$ can be reduced from the dynamical system up to some order by the coordinate transform.

There are several basic scales contained by the perturbation. The first one is the length scale of the normalized Larmor Radius $\varepsilon  \equiv \frac{{\rho} }{{{L_0}}}$ with ${\rho}  \equiv \frac{{m{v_t}}}{{{B_0}q}}$. The second one is the amplitude of the electrostatic potential, whose order is written as $O\left( {\left\| \phi  \right\|} \right) = O\left( {{\varepsilon ^\sigma }} \right)$, based on the basic parameter $\varepsilon$. In plasma, due to the fact that charged particle can nearly migrate freely in the environment with collective interactions, the magnitude of the potential the particles feel must be much smaller than that of its kinetic energy. As Eq.(\ref{a13}) shows, the order of the kinetic energy is $O(\varepsilon)$. Therefore, it's plausible to assume the range for $\sigma$ being $2\ge \sigma >1$. In Ref.(\onlinecite{1988hahm}), $\sigma=2$ is chosen. The third one is the length scale of the gradient of the electrostatic potential. Its order is recorded as $O(\varepsilon^{\beta})$. The case of $\beta<1$ is used in drift kinetic theory, while the case of $\beta \approx 1$ is for the gyrokinetic theory, which is the focus of our study.

\section{The conventional electrostatic gyrokinetic model}\label{sec3}

\subsection{The coordinate transform and the orbit equations}

The coordinate transform contained by CEGM is splitted into two independent steps. The first step is to carry out the guiding center transform, while the second step is to carry out the gyrocenter transform. If we replace $\varepsilon_2 L_{\mathbf{g}_2}$ by $L_{\mathbf{g}_2}$ and $\varepsilon_1 L_{\mathbf{g}_1}$ by $L_{\mathbf{g}_1}$ for the simplicity, the new Lagrangian differential 1-form on gyrocenter coordinate can be formally written as
\begin{equation}\label{g4}
\Gamma  = \exp \left( { - {L_{{{\bf{g}}_2}}}} \right)\exp \left(- {{L_{{{\bf{g}}_1}}}} \right)\gamma \left( {\bf{Z}} \right)
\end{equation}
where $\mathbf{g}_i\equiv(\mathbf{g}_i^\mathbf{X},g_i^\mu,g_i^U,g^{\theta}_i)$ for $i=1,2$ with $\mathbf{g}^\mathbf{X}_i\equiv (g_i^{1},g_i^2,g_i^3)$ being the spatial components. The subscript $i\in\{1,2\}$ are indexes for the guiding-center transform and gyrocenter transform, respectively. The guiding-center coordinate is denoted by $\bar{\mathbf{Z}}\equiv(\bar{\mathbf{X}},\bar{\mu},\bar{U},\bar{\theta})$, whilst ${\mathbf{Z}}\equiv({\mathbf{X}},{\mu},{U},{\theta})$ for the gyrocenter coordinate. It's well known \cite{1990brizard, 2007brizard1} that the generator for the guiding center is
\begin{equation}\label{g5}
{\bf{g}}_1^{\bf{X}} =  - \varepsilon {\bm{\rho} _0} =  - \varepsilon \sqrt {\frac{{2\mu }}{{B\left( {{{\bf{X}}_1}} \right)}}} \left( { - {{\bf{e}}_1}\cos \theta  + {{\bf{e}}_2}\sin \theta } \right).
\end{equation}
And the associated Lagrangian 1-form on guiding-center coordinates is
\begin{eqnarray}\label{g6}
\bar \Gamma  &&= \exp(-{L_{{{\bf{g}}_1^\mathbf{X}}}})\gamma \left( \bar{{\bf{Z}}} \right)  \nonumber  \\
&&= {\bf{A}}\left( {{\bf{\bar X}}} \right)\cdot d{\bf{\bar X}} + \varepsilon \bar U{\bf{b}}\cdot d{\bf{\bar X}} + {\varepsilon ^2}\bar \mu d\bar \theta   \nonumber \\
&& - \left[ {\varepsilon \left( {\frac{{\bar U_{}^2}}{2} + \bar \mu B\left( {{\bf{\bar X}}} \right)} \right) + \exp \left( {{\bm{\rho} _0}\left( {{\bf{\bar Z}}} \right)\cdot\nabla } \right)\phi \left( {{\bf{\bar X}},t} \right)} \right]dt.
\end{eqnarray}
Then, the new Lagrangian differential 1-form on gyrocenter coordinate $\Gamma  = \exp(-{L_{{{\bf{g}}_2}}}){\bar{\Gamma}}\left( {\bf{Z}} \right)$ is expanded. Among the expanding, the following terms are written as
\begin{equation}\label{g8}
{\Gamma _0} = {\bf{A}}\left( {\bf{X}} \right)\cdot d{\bf{X}} + \varepsilon U{\bf{b}}\cdot d{\bf{X}} + {\varepsilon ^2}\mu d\theta  -  \left( {\varepsilon \mu B\left( {\bf{X}} \right) + \varepsilon \frac{{m{U^2}}}{2}}+\phi(\mathbf{X},t) \right)dt.
\end{equation}
The terms linear to the generator $\mathbf{g}_2$ or depending on the perturbative potential are written together as
\begin{eqnarray}\label{g7}
{\Gamma _1} &=& \left( { - \left( {{\bf{B}} + \varepsilon U\nabla  \times {\bf{b}}} \right) \times {\bf{g}}_2^{\bf{X}} - \varepsilon g_2^U{\bf{b}}} \right)\cdot d{\bf{X}}  \nonumber \\
&& + \varepsilon \left( {{\bf{g}}_2^{\bf{X}}\cdot{\bf{b}}} \right)dU - {\varepsilon ^2}{g_2^\mu }d\theta  - {\varepsilon ^2}g_2^\theta d\mu  \nonumber \\
&& - \left( {\varepsilon \mu {\bf{g}}_2^{\bf{X}}\cdot\nabla B\left( {\bf{X}} \right) - \varepsilon Ug_2^U - \varepsilon g_2^\mu B + \left( {\exp \left( {\varepsilon {\bm{\rho} _0}\cdot{\nabla _{\bf{X}}}} \right) - 1} \right){\phi}} \right)dt \nonumber \\
&& + dS.
\end{eqnarray}
To get Eq.(\ref{g7}), the non-zero components of the Lie derivative on $\Gamma_0$ given by Sec.(\ref{app1}) are used.
To remove the $\theta$-dependent terms in Eq.(\ref{g7}), the following identities are required
\begin{equation}\label{g9}
\Gamma_{1k}=0,Z^k\in \{\mathbf{X},U,\mu,\theta\}
\end{equation}
plus a requirement that $\Gamma_{1t}$ is independent of $\theta$. Then, all the generators can be solved
\begin{equation}\label{g10}
{\bf{g}}_2^{\bf{X}} =  - \frac{{{\bf{b}} \times \nabla {S_1}}}{{{\bf{b}}\cdot{{\bf{B}}^*}}} - \frac{{{{\bf{B}}^*}}}{\varepsilon }\frac{{\partial {S_1}}}{{\partial U}},
\end{equation}
with
\begin{equation}\label{g11}
{{\bf{B}}^*} = {\bf{B}} + \varepsilon U\nabla  \times {\bf{b}},
\end{equation}
and
\begin{equation}\label{g12}
g_2^U = \frac{1}{\varepsilon }{\bf{b}}\cdot\nabla {S_1},
\end{equation}
\begin{equation}\label{g20}
g_2^\mu  = \frac{1}{{{\varepsilon ^2}}}\frac{{\partial {S_1}}}{{\partial \theta }},
\end{equation}
\begin{equation}\label{g13}
{g_2^\theta } =  - \frac{1}{{{\varepsilon ^2}}}\frac{{\partial {S_1}}}{{\partial \mu }}.
\end{equation}
The equation for the gauge function is
\begin{equation}\label{g14}
\frac{{\partial {S_1}}}{{\partial t}} + U{\bf{b}}\cdot\nabla {S_1} + \frac{1}{\varepsilon }\frac{{\partial {S_1}}}{{\partial \theta }} = \phi \left( {{\bf{X}} + \varepsilon {\bm{\rho} _0}} \right) + {\Gamma _{1t}}.
\end{equation}

For the low frequency perturbation, inequalities $\left| {\frac{{\partial {S_1}}}{{\partial t}}} \right| \ll \left| {\frac{B}{{{\varepsilon }}}\frac{{\partial {S_1}}}{{\partial \theta }}} \right|, \left| {U{\bf{b}}\cdot\nabla {S_1}} \right| \ll \left| {\frac{B}{{{\varepsilon }}}\frac{{\partial {S_1}}}{{\partial \theta }}} \right|$ hold, and the lowest order equation of Eq.(\ref{g14}) is
\begin{equation}\label{g15}
\frac{{B({\bf{X}})}}{\varepsilon }\frac{{\partial {S_1}}}{{\partial \theta }} = \phi ({\bf{X}} + \varepsilon {\bm{\rho} _0(\mathbf{Z})},t) + {\Gamma _{1t}}.
\end{equation}
To remove the secularity of $S_1$ on the integration of $\theta$, $\Gamma_{1t}$ is chosen as
\begin{equation}\label{g16}
{\Gamma _{1t}} =  - \left\langle {\exp \left( {\varepsilon {\bm{\rho} _0(\mathbf{Z})}\cdot\nabla } \right){\phi}\left( {{\bf{X}},t} \right)} \right\rangle,
\end{equation}
where the symbol $\left \langle  \right \rangle $ means the averaging on $\theta$. The reason for removing the secularity from $S_1$ is that those secular terms could contribute unlimited terms to the generators through Eqs. (\ref{g10},\ref{g12},\ref{g20},\ref{g13}), which causes the coordinate transform unacceptable.

 The new Lagrangian 1-form becomes
\begin{eqnarray}\label{g17}
\Gamma  &=& \left( {{\bf{A}}\left( {\bf{X}} \right) + \varepsilon U{\bf{b}}} \right)\cdot d{\bf{X}} + {\varepsilon ^2}\mu d\theta  \nonumber \\
&& - \left( {\varepsilon \left( {\mu B\left( {\bf{X}} \right)}  + \frac{{m{U^2}}}{2} \right)+ \left\langle {\exp \left( {\varepsilon {\bm{\rho} _0}\cdot\nabla } \right)\phi \left( {\bf{X}} \right)} \right\rangle } \right)dt.
\end{eqnarray}
Applying the variational principle to this Lagrangian 1-form given by Eq.(\ref{g17}), the orbit equations are derived as
\begin{equation}\label{g18}
\mathop {\bf{X}}\limits^. {\rm{ = }}\frac{{U{{\bf{B}}^*} + {\bf{b}} \times \nabla \left( {\mu B\left( {\bf{X}} \right) + \left\langle {\exp \left( {\varepsilon {\bm{\rho} _0}\cdot\nabla } \right)\phi \left( {\bf{X}} \right)} \right\rangle } \right)}}{{{\bf{b}}\cdot{{\bf{B}}^*}}},
\end{equation}
\begin{equation}\label{g19}
\dot U = \frac{{ - {{\bf{B}}^*}\cdot\nabla \left( {\mu B\left( {\bf{X}} \right) + \left\langle {\exp \left( {\varepsilon {\bm{\rho} _0}\cdot\nabla } \right)\phi \left( {\bf{X}} \right)} \right\rangle } \right)}}{{\varepsilon {\bf{b}}\cdot{{\bf{B}}^*}}},
\end{equation}
with $\dot{\mu}=0$.

\subsection{Derive CEGM}

The principle of gyrokinetic model is that through the coordinate transform, the distribution function contained by the Vlasov equation is on gyrocenter coordinate and independent of the gyrophase for the reduction of the simulation burden, while the Poisson equation is still on the particle coordinates. The Vlasov equation is
\begin{equation}\label{a46}
\left( {\frac{\partial }{{\partial t}} + \frac{{d{\bf{X}}}}{{dt}}\cdot\nabla  + \frac{d}{{dU}}\frac{\partial }{{\partial U}}} \right)F_s\left( {{\bf{Z}},t} \right) = 0,
\end{equation}
where the subscript $s$ denotes the species of charged particles.
It needs two independent steps to transform the distribution function on gyrocenter coordinates to the one on particle coordinates, denoted by the following formula
\begin{equation}\label{gg20}
F_s\left( {{\bf{X}},\mu ,U},t \right)\mathop{\longrightarrow} \limits^{\psi _{gy}^{ - 1}} \bar{F}_s\left(\bar{\mathbf{Z}} ,t \right)\mathop{\longrightarrow} \limits^{\psi _g^{ - 1}} f_s\left(\mathbf{z},t \right)
\end{equation}
where $\psi_{gy}^{-1}$ and $\psi_{g}^{-1}$ denote the reverse gyrocenter transform and reverse guiding-center transform, respectively, and $\mathbf{z}\equiv (\mathbf{x},\mu_1,u_1,\theta_1)$.  To get the distribution on particle coordinates, on one hand, the  total distribution function is separated into the sum of an equilibrium one and a perturbative one as
\begin{equation}\label{21}
F_s\left( {{\bf{Z}},t} \right) = {F_{s0}}\left( {{\bf{Z}}} \right) + {F_{s1}}\left( {{\bf{Z}},t} \right).
\end{equation}
 On the other hand, the order of the magnitude of $g_2^\theta$ and $g_2^\mu$ is lower than that of $g_2^U$ and $\mathbf{g}_2^\mathbf{X}$, so the coordinate transform $\psi _{gy}^{ - 1}:{\bf{Z}} \to \overline {\bf{Z}}$ is linearly approximated as
\begin{equation}\label{g22}
{\bf{\bar X}} = {\bf{X}},\bar \mu  = \mu  - g_2^\mu \left( {\bar {\bf{Z}} } \right),\bar U = U,\bar \theta  = \theta  - g_2^\theta \left( {\bar {\bf{Z}} } \right),
\end{equation}
linear to the order of the normalized amplitude of the perturbative wave.
By substituting the coordinate transform into $F(\mathbf{Z})$ , the linear approximation of the distribution function on guiding-center coordinates is
\begin{equation}\label{g23}
{{\bar F}_s}\left( {{\bf{\bar X}},\bar \mu ,\bar U,\bar \theta } \right) = g_2^\mu \left( {\bar {\bf{Z}} } \right){\partial _{\bar \mu }}{{\bar F}_{s0}}\left( {{\bf{\bar X}},\bar \mu ,\bar U} \right) + {{\bar F}_{s1}}\left( {\bar {\bf{Z}} } \right).
\end{equation}
where the units of $g_2^\mu(\bar{\mathbf{Z}})$ is recovered with the result being
\begin{equation}\label{g25}
g_2^\mu \left( {\overline {\bf{Z}} } \right) = \frac{{{q_s}}}{{B\left( {\overline {\bf{X}} } \right)}}\left[ {\phi \left( {\overline {\bf{X}}  + {\bm{\rho} _0},t} \right) - \left\langle {\exp \left( {\varepsilon {\bm{\rho} _0}\cdot\nabla } \right)\phi } \right\rangle \left( {\overline {\bf{X}} ,t} \right)} \right].
\end{equation}

The reverse guiding-center coordinate transform $\psi _g^{ - 1}:\bar {\bf{Z}}  \to {\bf{z}}$ is linearly approximated as
\begin{equation}\label{g24}
\emph{}{\bf{x}} = {\bf{\bar X}} + {\bm{\rho} _0},{\mu _1} = \bar \mu ,{u_1} = \bar U,{\theta _1} = \bar \theta.
\end{equation}
Then, the transformation of the distribution function from guiding-center coordinates to the particle coordinates is approximated as
\begin{eqnarray}\label{g26}
f\left( {\bf{z}} \right) &=& \int \begin{array}{l}
{{\bar F}_s}\left( {\overline {\bf{Z}} } \right)\delta \left( {{\bf{x}} - {\bf{\bar X}} - {\bm{\rho} _0}\left( {\overline {\bf{Z}} } \right)} \right)\delta \left( {{\mu _1} - \bar{\mu}+g_2^\mu \left( {\overline {\bf{Z}} } \right) } \right) \\
\delta \left( {{u_1} - \bar U} \right)\delta \left( {{\theta _1} - \bar \theta } \right)J{d^3}\overline {\bf{X}} d\bar \mu d\bar Ud\bar \theta
\end{array}  \nonumber \\
 &\approx& F_{s0}(\mathbf{z}) - \frac{{{q_s}}}{{{T_s }}}\left[ {\phi \left( {{\bf{x}},t} \right)- \left\langle {\exp \left( {{\bm{\rho} _0}(\mathbf{z})\cdot\nabla } \right)\phi } \right\rangle \left( {{\bf{x}} - {\bm{\rho} _0}\left( {\bf{z}} \right),t} \right)} \right] {F_{s0}}\left( {\bf{z}} \right)   \nonumber \\
&& + {{\bar F}_{s1}}\left( {{\bf{x}} - {\bm{\rho} _0}\left( {\bf{z}} \right),{\mu _1},{u_1},{\theta _1}} \right),
\end{eqnarray}
where the Jacobian of $\psi_g^{-1}$ is approximated to equal one.
The density can be obtained by integrating $f\left( {\bf{z}} \right)$ on velocity space
\begin{eqnarray}\label{g27}
{n_s}\left( {{\bf{x}},t} \right) &=& {n_{s0}}({\bf{x}}) - \frac{{{q_s}}}{{{T_s}}}\left[ \begin{array}{l}
{n_{s0}}\left( {\bf{x}} \right)\phi ({\bf{x}},t)\\
 - \left\langle {\left\langle {\exp \left( {{\bm{\rho} _0}(\mathbf{z})\cdot\nabla } \right)\phi } \right\rangle \left( {{\bf{x}} - {\bm{\rho} _0}\left( {\bf{z}} \right),t} \right)} \right\rangle
\end{array} \right] \nonumber \\
&& + {n_{s1}}\left( {{\bf{x}},t} \right)
\end{eqnarray}
with
\begin{eqnarray}\label{g28}
&& \left\langle {\left\langle {\exp \left( {{\bm{\rho} _0(\mathbf{z})}\cdot\nabla } \right)\phi } \right\rangle \left( {{\bf{x}} - {\bm{\rho} _0}\left( {\bf{z}} \right),t} \right)} \right\rangle  \nonumber \\
&& \equiv \int {\left\langle {\exp \left( {{\bm{\rho} _0}(\mathbf{z})\cdot\nabla } \right)\phi } \right\rangle \left( {{\bf{x}} - {\bm{\rho} _0(\mathbf{z})}\left( {\bf{z}} \right),t} \right){F_{s0}}\left( {\bf{z}} \right)\frac{B(\mathbf{x})}{{{m_s}}}d{\mu _1}d{u_1}d{\theta _1}}  \nonumber \\
&& = \int {\exp \left( { - {\bm{\rho} _0(\mathbf{z})}\cdot\nabla } \right)\left\langle {\exp \left( {{\bm{\rho} _0(\mathbf{z})}\cdot\nabla } \right)\phi } \right\rangle \left( {{\bf{x}},t} \right){F_{s0}}\left( {\bf{z}} \right)\frac{B(\mathbf{x})}{{{m_s}}}d{\mu _1}d{u_1}d{\theta _1}}
\end{eqnarray}
and
\begin{equation}\label{g31}
{n_{s1}}\left( {{\bf{x}},t} \right) \equiv \int {{{\bar F}_{s1}}\left( {{\bf{x}} - {\bm{\rho} _0}\left( {\bf{z}} \right),{\mu _1},{u_1},{\theta _1}} \right)\frac{{B\left( {\bf{x}} \right)}}{{{m_s}}}d{\mu _1}d{u_1}d{\theta _1}}.
\end{equation}
Here, $\frac{B(\mathbf{x})}{m_s}$ is the Jacobian of the coordinate transform between $\mathbf{v}$ in Cartesian coordinate and $(\mu_1,u_1,\theta_1)$ in cylindrical coordinate.
The quasi-neutral equation can be derived as
\begin{equation}\label{g29}
\sum\limits_s {{q_s}\left[ { - \frac{{{q_s}}}{{{T_s}}}\left[ {{n_{s0}}\left( {\bf{x}} \right)\phi \left( {{\bf{x}},t} \right) - \left\langle {\left\langle {\exp \left( {{\bm{\rho} _0}(\mathbf{z})\cdot\nabla } \right)\phi } \right\rangle \left( {{\bf{x}} - {\bm{\rho} _0}\left( {\bf{z}} \right),t} \right)} \right\rangle } \right] + {n_{s1}}} \right]}  = 0.
\end{equation}
If the plasmas only include proton and electron and the distribution of electrons is approximated by the adiabatic distribution,
the quasi-neutral equation is simplified as
\begin{equation}\label{g30}
\frac{e}{{{T_i}}}\left\langle {\left\langle {\exp \left( {{\bm{\rho} _0}(\mathbf{z})\cdot\nabla } \right)\phi } \right\rangle \left( {{\bf{x}} - {\bm{\rho} _0}\left( {\bf{z}} \right),t} \right)} \right\rangle  - \frac{e}{{{T_i}}}{n_{0}}\phi \left( {{\bf{x}},t} \right) + {n_{i1}} - \frac{e}{{{T_e}}}{n_{0}}\phi \left( {{\bf{x}},t} \right) = 0.
\end{equation}

Here, it's necessary to emphasize on an obvious difference between the model derived here and the usual CEGM. In  Eq.(\ref{g26}), the terms associated with the potential function is given as ${\phi \left( {{\bf{x}},t} \right) - \left\langle {\exp \left( {{\bm{\rho} _0}\cdot\nabla } \right)\phi } \right\rangle \left( {{\bf{x}} - {\bm{\rho} _0}\left( {\bf{z}} \right),t} \right)}$, while in the usual conventional gyrokinetic model, it's given as $\phi \left( {{\bf{x}},t} \right) - \left\langle {\exp \left( {{\bm{\rho} _0}\cdot\nabla } \right)\phi } \right\rangle \left( {{\bf{x}},t} \right)$. The latter is not right, which can be recognized in terms of Eqs.(\ref{g25},\ref{g26}).

Now the derivation of CEGM is completed, which comprises Eqs.(\ref{g18}),(\ref{g19}),(\ref{a46}) and (\ref{g29}).

\section{Electrostatic gyrokinetic model derived through the covariant transform formula}\label{sec5}

\subsection{Simple introduction of the covariant transform formula of the differential one-form}\label{sec4}

To begin, it's first assumed that $\psi$ is a general coordinate transformation defined as ${\psi }:{\bf{Y}} \to {\bf{y}}$ , where $\mathbf{Y}$ and $\mathbf{y}$ are both a $p$-dimensional manifold. $\gamma \left( {\bf{y}} \right) \equiv {\gamma _k}\left( {\bf{y}} \right)d{y^k}$ is a differential 1-form defined on $\bf{y}$, where $k\in\{1,\cdots,p\}$ and the repeated indexes means the summation of all $k$s. $\mathbf{v}_k$ is a tangent vector defined on $\mathbf{Y}$. Here, $\mathbf{v}_k$ is chosen to be $\mathbf{v}_k=\partial/\partial Y_k$. $\psi$ induces a pullback transformation ${\psi^*}$ for $\gamma$, with the new $\Gamma$ written as $\Gamma=\psi^* \gamma$ and defined on $\mathbf{Y}$.  The pullback transform of $\gamma$ is defined as
\begin{equation}\label{a1}
{\left. {\left( {{\psi ^*}\gamma(\mathbf{y},\varepsilon) } \right)} \right|_{\bf{Y}}}\left( {\frac{\partial }{{\partial {Y^k}}}} \right) = {\left. \gamma(\mathbf{y},\varepsilon)  \right|_{\bf{y}}}\left( {{{\left. {d\psi \left( {\frac{\partial }{{\partial {Y^k}}}} \right)} \right|}_{\bf{y}}}} \right),
\end{equation}
based on the contraction rule between the tangent vector and the cotangent vector \cite{2006marsden, 1989arnoldbook}.
The formula ${d\psi \left( {\frac{\partial }{{\partial {Y^k}}}} \right)}$ is a pushforward transformation of $\partial/\partial Y^k$ with the definition $d\psi \left( {\frac{\partial }{{\partial {Y^k}}}} \right) \equiv \frac{{\partial {\psi ^k}\left( {\bf{Y}} \right)}}{{\partial {Y^k}}}\frac{\partial }{{\partial {y^k}}}$. Substituting the pushforward transformation to Eq.(\ref{a1}) and adopting the contraction rule, the $i$th component of new 1-form with $i\in\{1,\cdots,p\}$ transformed from $\gamma$  is
\begin{equation}\label{a2}
{\Gamma _i}({\bf{Y}},\varepsilon ) = {\gamma _k}(\psi ({\bf{Y}}))\frac{{\partial {\psi ^k}\left( {\bf{Y}} \right)}}{{\partial {Y^i}}},
\end{equation}
which is also called the covariant transform formula of the 1-form as the covariant vector.

\subsection{Simple introduction of the expanding of the pullback formula}
Eq.(\ref{a2}) associates the original differential one-form, the new one-form and the coordinate transform together. By applying Eq.(\ref{a2}) to our example, we need to make the following replacement: $\mathbf{Y}\to \mathbf{Z}$; $\mathbf{y}\to\mathbf{z}$. $\gamma$ in Eq.(\ref{a2}) is given by Eq.(\ref{a11}) and $\Gamma$ is the new 1-form to be solved. For convenience, the coordinate transform $\mathbf{y}=\psi(\mathbf{Y})$ is replaced by another symbol as
\begin{equation}\label{gf1}
\mathbf{z}\equiv \mathbf{Z}_b(\mathbf{Z},E_2),
\end{equation}
where $E_2\equiv\{\varepsilon,\varepsilon^\sigma\}$ is the set of the two parameters denoting the scales contained by the perturbation. The formula for the pullback transform can also be rewritten as
\begin{equation}\label{gf2}
{\Gamma _h}\left( {{\bf{Z}},{E_2}} \right) = \frac{{\partial Z_b^k}}{{\partial Z_{}^h}}\left( {{\bf{Z}},{E_2}} \right){\gamma _k}\left( {{{\bf{Z}}_b}\left( {{\bf{Z}},{E_2}} \right),{E_2}} \right)
\end{equation}
with $Z^h,Z^k\in\{\mathbf{X},\mu,U,\theta\}$.

${{{\bf{Z}}_b}\left( {{\bf{Z}},{E_2}} \right)}$ can be formally expanded based on the basic parameter set
${E}_2$
\begin{equation}\label{gf3}
Z_b^k ({\bf{Z}},{{E}}_2) = {Z^k} + Z_b^{k *}({\bf{Z}},{{E}}_2),
\end{equation}
with
\begin{equation}\label{gf4}
Z_b^{k*}({\bf{Z}},{E_2}) = {\varepsilon ^m}Z_{b,m}^k({\bf{Z}}),
\end{equation}
where the repeated $m$ means the summation of the all the indexes; the case of $m=0$ is deleted from Eq.(\ref{gf4}); the order of $Z_{b,m}^k({\bf{Z}})$ is $O(1)$. However, all $\varepsilon^m$s and $Z_{b,m}^k({\bf{Z}})$s are unknown, needing to be solved under the purpose of reducing the gyrophase from the new Lagrangian 1-form.

The formula of $\gamma(\mathbf{Z}_b(\mathbf{Z},{E}_2),{E}_2)$ can also be expanded. It's first written as the sum
\begin{equation}\label{gf5}
\gamma ({{\bf{Z}}_b}({\bf{Z}},{E_2}),{E_2}) = \left( {{\gamma _0} + {\gamma _1} + {\gamma _\sigma }} \right)\left( {{{\bf{Z}}_b}({\bf{Z}},{E_2})} \right)
\end{equation}
where $\gamma_0,\gamma_1,\gamma_\sigma$ are given in Eqs.(\ref{a12},\ref{a13},\ref{a14}), respectively. Then, $\gamma ({{\bf{Z}}_b}({\bf{Z}},{E_2}),{E_2})$ can be expanded as
\begin{equation}\label{gf6}
\gamma ({{\bf{Z}}_b}({\bf{Z}},{E_2}),{E_2}) = \frac{{{{\left( {{\bf{Z}}_b^*({\bf{Z}},{E_2})} \right)}^n}}}{{n!}}:{\left( {\frac{\partial }{{\partial {\bf{Z}}}}} \right)^n}\left( {{\gamma _0} + {\gamma _1} + {\gamma _\sigma }} \right)\left( {\bf{Z}} \right),
\end{equation}
where symbol `$:$' means the inner product between two tensors and $k\ge 0$.

To cancel those perturbative terms depending on fast variables at a certain order, we  introduce new  and appropriate $\varepsilon^m$ and $\mathbf{Z}_{b,m}^*$ into the coordinate transform. By introducing this term, there would be new terms generated in the expanding in Eq.(\ref{gf2}). The lowest order terms of which are designed to cancel the already existed terms depending on the gyrophase at that order. By repeatedly introducing new $\varepsilon^m$ and $\mathbf{Z}_{b,m}^*$ , the order of the terms depending on the gyrophase in the new $\Gamma$ becomes higher and higher.

\subsection{The specific procedure to cancel the $\theta$-dependent terms in the new Lagrangian 1-form}

The zero order of the expansion in Eq.(\ref{gf2}) is trivial and given as $\mathbf{A}(\mathbf{X})$. The term depending on $\theta$ and possessing the order $O(\varepsilon)$ is $\gamma_{1\mathbf{X}\perp}\cdot d\mathbf{X}$ denoted as $\Gamma_{1o}(\theta)$, where  the subscript `$o$' denotes the term inherited from the original 1-form.  To cancel this term, we introduce the first coordinate-transform factor denoted as $\varepsilon\mathbf{Z}_{b,1}^\mathbf{X}$ only in the spatial component, as the superscript $\mathbf{X}$ indicates. Then, the lowest order terms generated by the introduction of $\mathbf{Z}_{b,1}^\mathbf{X}$ are as follows
\begin{eqnarray}\label{a49}
{\Gamma _1^{'}} &=& \varepsilon\left( {\frac{{\partial Z_{b,1}^{Xl}}}{{\partial {X^c}}}{\gamma _{0l}} + Z_{b,1}^{Xl}\frac{\partial }{{\partial {X^l}}}{\gamma _{0c}}} \right)d{X^c} + \varepsilon\sum\limits_{Z^j \in \{ \mu ,U,\theta \} } {\frac{{\partial Z_{b,1}^{Xl}}}{{\partial {Z^j}}}{\gamma _{0l}}}dZ^j \nonumber \\
 &=& \varepsilon Z_{b,1}^{Xl}\left( {\frac{{\partial {\gamma _{0c}}}}{{\partial {X^l}}} - \frac{{\partial {\gamma _{0l}}}}{{\partial {X^c}}}} \right)d{X^c} + \varepsilon\frac{\partial }{{\partial {Z^k}}}\left( {Z_{b,1}^{Xl}{\gamma _{0l}}} \right)d{Z^k} \nonumber \\
 &=&  - \varepsilon{\bf{Z}}_{b.1}^{\bf{X}} \times {\bf{B}}\left( {\bf{X}} \right)\cdot d{\bf{X}} + dS
\end{eqnarray}
where $X^l,X^c\in\{X^1,X^2,X^3\}$ and $Z^{Xl}_{b,1}$ for $l\in\{1,2,3\}$ is the $X^l$ component of $\mathbf{Z}^{\mathbf{X}}_{b,1}$. In Eq.(\ref{a49}), the equalities $\frac{{\partial {\gamma _{0l}}}}{{\partial {Z^j}}}=0$ for $Z^j \in \{ \mu ,U,\theta \}$ are used.
The non-gauge term in Eq.(\ref{a49}) is applied to cancelling $\Gamma_{1o}(\theta)$. The solution of $\mathbf{Z}_{b,1}^\mathbf{X}$ is
\begin{equation}\label{a50}
\mathbf{Z}_{b,1}^\mathbf{X}={\bm{\rho} _0}=\sqrt {\frac{{2\mu }}{{B\left( {\bf{X}} \right)}}} \left( { - {{\bf{e}}_1}\cos \theta  + {{\bf{e}}_2}\sin \theta } \right).
\end{equation}

The introduction of ${\bf{Z}}_{b.1}^{\bf{X}}$ generates the following $\Gamma_\sigma^{'}(\theta)$
\begin{equation}\label{a51}
{\Gamma _\sigma^{'} }\left( \theta  \right) =  - \left[ {\exp \left( {\varepsilon {\bm{\rho} _0}\cdot\nabla_\sigma } \right) - 1} \right]\phi \left( {{\bf{X}},t} \right)dt.
\end{equation}
This term possesses the order $O(\varepsilon^\sigma)$, since $O(\left\| {{\nabla _\sigma }} \right\|) = O\left( \varepsilon^{-1}  \right)$ holds for $\nabla_\sigma$ only acting upon $\phi(\mathbf{x},t)$. To cancel this term, we introduce the second factor $\varepsilon^{\sigma-1}Z^{\mu}_{b,\sigma-1}$ only including the $\mu$ component, the lowest order terms generated by which possess the order $O(\varepsilon^\sigma)$ and are given as follows
\begin{equation}\label{a52}
\Gamma _{\sigma 1} ^{'} = {\varepsilon ^{\sigma-1} }Z_{b,\sigma  - 1}^\mu {\partial _\mu }{\gamma _{1{\bf{X}} \bot }}\cdot d{\bf{X}} - {\varepsilon ^\sigma }Z_{b,\sigma  - 1}^\mu B\left( {\bf{X}} \right)dt,
\end{equation}
where $\gamma_{1\mathbf{X}\perp}$ possesses the order $O(\varepsilon)$.
The time component in Eq.(\ref{a52}) is used to cancel ${\Gamma _{\sigma }^{'}}(\theta)$. Then, it can be solved that
\begin{equation}\label{a53}
{\varepsilon ^{\sigma  - 1}}Z_{b,\sigma  - 1}^\mu  =  - \frac{{\left[ {\exp \left( {\varepsilon {\bm{\rho} _0}\cdot\nabla_\sigma } \right) - 1} \right]\phi \left( {{\bf{X}},t} \right)}}{{\varepsilon B\left( {\bf{X}} \right)}}.
\end{equation}
The spatial components in Eq.(\ref{a52}) still need to be cancelled, which is completed by introducing the third factor $\varepsilon^\sigma \mathbf{Z}_{b,\sigma}^\mathbf{X}$. The lowest order generated by this term is $O(\varepsilon^{\sigma})$ with the associated terms being
\begin{eqnarray}\label{a54}
{\Gamma _{\sigma 2}^{'}} &=& \varepsilon^\sigma \left( {\frac{{\partial Z_{b,\sigma }^{Xl}}}{{\partial {X^c}}}{\gamma _{0l}} + Z_{b,\sigma }^{Xl}\frac{{\partial {\gamma _{0c}}}}{{\partial {X^l}}}} \right)d{X^c} + {\varepsilon ^{\sigma }}\frac{{\partial Z_{b,\sigma }^{Xl}}}{{\partial {X^c}}}{\gamma _{1{\bf{X}}l}}d{X^c} \nonumber \\
 &=&  - \varepsilon^\sigma {\bf{Z}}_{b,\sigma }^{\bf{X}} \times {\bf{B}}\left( {\bf{X}} \right)\cdot d{\bf{X}}  + {\varepsilon ^{\sigma }} {\frac{{\partial Z_{b,\sigma }^{Xl}}}{{\partial {X^c}}}{\gamma _{1{\bf{X}}l}}} d{X^c}+ dS.
\end{eqnarray}
The reason for the presentation of the second term in the first equality of Eq.(\ref{a54}) is that $\gamma_{1\mathbf{X}\perp}$ possesses the order $O(\varepsilon)$ and $O\left( {\left\| {\frac{\partial }{{\partial {X^l}}}} \right\|} \right) = O\left( {{\varepsilon ^{ - 1}}} \right)$ for $\frac{\partial}{\partial X^l}$ acting upon $\phi(\mathbf{X},t)$ contained by $Z_{b,\sigma}^{Xl}$. The first term of the second equality in Eq.(\ref{a54}) is used to canceling the spatial components in Eq.(\ref{a52}), which leads to the solution as
\begin{equation}\label{a55}
{\bf{Z}}_{b,\sigma} ^{\bf{X}} = \frac{{ - 1}}{\varepsilon {B\left( {\bf{X}} \right)}}Z_{b,\sigma  - 1}^\mu {\partial _\mu }{\gamma _{1{\bf{X}} \bot }} \times {\bf{b}}.
\end{equation}
On the other hand, the second term in Eq.(\ref{a54}) can be rewritten as
\begin{eqnarray}\label{a56}
{\varepsilon ^{\sigma }}\frac{{\partial Z_{b,\sigma }^{Xl}}}{{\partial {X^c}}}{\gamma _{1{\bf{X}}l}}d{X^c} &=& {\varepsilon ^{\sigma }}\left[ {\frac{{\partial \left( {{\bf{Z}}_{b,\sigma }^{\bf{X}}\cdot{\gamma _{1{\bf{X}}}}} \right)}}{{\partial {X^c}}} - Z_{b,\sigma }^{Xl}\frac{{\partial {\gamma _{1{\bf{X}}l}}}}{{\partial {X^c}}}} \right]d{X^l} \nonumber \\
 &=&  - {\varepsilon ^{\sigma}}Z_{b,\sigma }^{Xl}\frac{{\partial {\gamma _{1{\bf{X}}l}}}}{{\partial {X^c}}}d{X^c},
\end{eqnarray}
since ${{\bf{Z}}_{b,\sigma }^{\bf{X}}\cdot{\gamma _{1{\bf{X}}}}}=0$. This term in fact possesses the order $O(\varepsilon^{\sigma+1})$ and is ignored, since we only keep terms possessing the order lower than or equaling the order of the amplitude of $\phi(\mathbf{X},t)$.

Eventually, after introducing $\varepsilon \mathbf{Z}_{b,1}^\mathbf{X}$, $\varepsilon^{\sigma-1}Z_{b,\sigma  - 1}^\mu $ and $\varepsilon^\sigma {\bf{Z}}_{b,\sigma }^{\bf{X}}$, the new Lagrangian 1-form, which is approximated up to the order $O(\varepsilon^\sigma)$ equaling the order of the amplitude of $\phi(\mathbf{X},t)$ is
\begin{equation}\label{a27}
\Gamma  = \left( {{\bf{A}}\left( {\bf{X}} \right) + \varepsilon U{\bf{b}}} \right)\cdot d{\bf{X}} - \left[ {\varepsilon \left( {\frac{{{U^2}}}{2} + \mu B({\bf{X}})} \right) + \phi ({\bf{X}},t)} \right]dt.
\end{equation}
An obvious difference from the conventional gyrokinetic theory is that FLR terms don't exist in the new differential 1-form.

The coordinate transform is
\begin{subequations}\label{a57}
\begin{eqnarray}
{\bf{x}} &=& {\bf{X}} + \varepsilon {\bm{\rho} _0}(\mathbf{Z}) + {\varepsilon ^\sigma }{\bf{Z}}_{b,\sigma }^{\bf{X}}(\mathbf{Z}),\;\;\;\\
{\mu _1} &=& \mu  + {\varepsilon ^{\sigma  - 1}}Z_{b,\sigma  - 1}^\mu(\mathbf{Z}) ,\\
{u_1} &=& U,\\
{\theta _1} &=& \theta
\end{eqnarray}
\end{subequations}

By imposing the variational principle on the 1-form, the orbit equations can be derived as
\begin{equation}\label{a28}
\mathop {\bf{X}}\limits^. {\rm{ = }}\frac{{U{{\bf{B}}^*} - {\bf{b}} \times \nabla \left( {\varepsilon \mu B\left( {\bf{X}} \right) + \phi ({\bf{X}},t)} \right)}}{{{\bf{b}}\cdot{{\bf{B}}^*}}},
\end{equation}
\begin{equation}\label{a29}
\dot U = \frac{{{{\bf{B}}^*}\cdot\nabla \left( {\varepsilon \mu B\left( {\bf{X}} \right) + \phi ({\bf{X}},t)} \right)}}{{\varepsilon {\bf{b}}\cdot{{\bf{B}}^*}}},
\end{equation}
with ${{\bf{B}}^*}\left( {\bf{X}} \right) = {\bf{B}}\left( {\bf{X}} \right) + \varepsilon U\nabla  \times {\bf{b}}$.

\subsection{The new electrostatic gyrokinetic model}

The units of all physical quantities are recovered first. By doing so, the orbit equations with units are
\begin{equation}\label{a32}
\mathop {\bf{X}}\limits^. {\rm{ = }}\frac{{U{{\bf{B}}^*} - {\bf{b}} \times \nabla \left( {\mu B\left( {\bf{X}} \right) + q\phi ({\bf{X}},t)} \right)}}{{{\bf{b}}\cdot{{\bf{B}}^*}}},
\end{equation}
\begin{equation}\label{a33}
\dot U = \frac{{{{\bf{B}}^*}\cdot\nabla \left( {\mu B\left( {\bf{X}} \right) + q\phi ({\bf{X}},t)} \right)}}{{m{\bf{b}}\cdot{{\bf{B}}^*}}},
\end{equation}
with ${{\bf{B}}^*} = {\bf{B}} + \frac{{mU}}{q}\nabla  \times {\bf{b}}$. The two formulas for $\varepsilon \bm{\rho}_0$ and ${Z}^\mu_{b,\sigma-1}$ after recovering the units are
\begin{equation}\label{a58}
\varepsilon \bm{\rho}_0=\frac{1}{q}\sqrt {\frac{{2m{\mu}}}{{B\left( {\bf{X}} \right)}}} \left( { - {{\bf{e}}_1}\cos \theta  + {{\bf{e}}_2}\sin \theta } \right),
\end{equation}
\begin{equation}\label{a35}
\varepsilon^{\sigma-1} Z_{b,\sigma  - 1}^\mu  =  - \frac{{q\left[ {\exp \left( { {\bm{\rho} _0}\cdot\nabla_\sigma } \right) - 1} \right]\phi \left( {{\bf{X}},t} \right)}}{{ B\left( {\bf{X}} \right)}}..
\end{equation}
The factor $\mathbf{Z}^\mathbf{X}_{b,\sigma}$ is ignored in the coordinate transform of the distribution, since it generates higher order terms.

For a plasma containing more than one species of particles,  the Vlasov equation is
\begin{equation}\label{a34}
\left( {\frac{\partial }{{\partial t}} + \frac{{d{\bf{X}}}}{{dt}}\cdot\nabla  + \frac{d}{{dU}}\frac{\partial }{{\partial U}}} \right)F_s\left( {{\bf{Z}},t} \right) = 0.
\end{equation}
The distribution function ${F_s\left( {{\bf{Z}},t} \right)}$ is on gyrocenter coordinates and subscript $s$ denotes the species. The distribution function on the particle coordinates is derived by the following transformation
\begin{equation}\label{a35}
f_s\left(\mathbf{z},t \right) = \int \begin{array}{l}
F_s\left( {{\bf{Z}},t} \right)\delta \left( {{\bf{x}} - {\bf{X}} - {\bm{\rho} _0}}(\mathbf{Z}) \right)\delta \left( {{\mu _1} - \mu  - \varepsilon^{\sigma-1} Z_{b,\sigma  - 1}^\mu(\mathbf{Z}) } \right)\\
\delta \left( {{u_1} - U} \right)\delta \left( {{\theta _1} - \theta } \right){d^3}{\bf{X}}d\mu dU{{d\theta }}.
\end{array}
\end{equation}
The distribution $F_s(\mathbf{Z},t)$ can be decomposed as an equilibrium Maxwellian one and a perturbative one
\begin{equation}\label{a36}
F_s\left( {{\bf{Z}},t} \right) = {F_{s0}}\left( {{\bf{Z}}} \right) + {F_{s1}}\left( {{\bf{Z}},t} \right).
\end{equation}
The equilibrium Maxwellian distribution function is
\begin{equation}\label{a39}
{F_{s0}}\left( {\bf{Z}} \right) \equiv {n_{s0}}\left( {\bf{X}} \right){\left( {\frac{m_s}{{2\pi {T_{s}}\left( {\bf{X}} \right)}}} \right)^{3/2}}\exp \left( {\frac{{ - m_s{U^2} - \mu B\left( {\bf{X}} \right)}}{{2{T_{s}}\left( {\bf{X}} \right)}}} \right).
\end{equation}
Then, expanding the integral in Eq.(\ref{a35}) and ignoring high order terms, the distribution function on particle coordinates can be approximated as
\begin{eqnarray}\label{a37}
{f_s}\left( {{\bf{z}},t} \right)
 &=& {F_{s0}}({\bf{z}}) - \frac{{{q_s}}}{{{T_s}}}\left( {1 - \exp \left( { - {\bm{\rho} _0}\left( {\bf{z}} \right)\cdot{\nabla _\sigma }} \right)} \right)\phi \left( {{\bf{x}},t} \right){F_{s0}}\left( {\bf{z}} \right) \nonumber \\
 && + {F_{s1}}\left( {{\bf{x}} - {\bm{\rho} _0}\left( {\bf{z}} \right),{\mu _1},{u_1},{\theta _1},t} \right),
\end{eqnarray}
linear to the amplitude of the perterbative wave.
To get Eq.(\ref{a37}), the following formula is used
\begin{eqnarray}\label{a38}
&& - {F_{s0}}({\bf{z}}) + \int \begin{array}{*{20}{l}}
{{F_{s0}}\left( {\bf{Z}} \right)\delta \left( {{\bf{x}} - {\bf{X}} - {\bm{\rho} _0}(\mathbf{Z})} \right)\delta \left( {{\mu _1} - \mu  - {\varepsilon ^{\sigma  - 1}}Z_{b,\sigma  - 1}^\mu }(\mathbf{Z}) \right)}\\
{\delta \left( {{u_1} - U} \right)\delta \left( {{\theta _1} - \theta } \right){d^3}{\bf{X}}d\mu dUd\theta }
\end{array}  \nonumber \\
 &\approx & \int \begin{array}{l}
 - {\varepsilon ^{\sigma  - 1}}Z_{b,\sigma  - 1}^\mu ({\bf{X}},{\mu _1},{u_1},{\theta _1}){\partial _\mu }{F_{s0}}\left( {{\bf{X}},{\mu _1},{u_1}} \right)\\
\delta \left( {{\bf{x}} - {\bf{X}} - {\bm{\rho} _0}} \right){d^3}{\bf{X}}
\end{array}  \nonumber \\
 &=&  - \frac{{{q_s}\exp {{\left( { - {\bm{\rho} _0}\cdot{\nabla _\sigma }} \right)}^*}\left[ {\exp \left( {{\bm{\rho} _0}\cdot{\nabla _\sigma }} \right) - 1} \right]\phi ({\bf{x}},t)}}{{{T_s}}}{F_0}({\bf{z}})  \nonumber \\
 &=&  - \frac{{{q_s}}}{{{T_s}}}\left( {1 - \exp \left( { - {\bm{\rho} _0}\left( {\bf{z}} \right)\cdot{\nabla _\sigma }} \right)} \right)\phi \left( {{\bf{x}},t} \right){F_{s0}}\left( {\bf{z}} \right).
\end{eqnarray}
 The term $\exp {\left( { - {\bm{\rho} _0}\left( {\bf{z}} \right)\cdot{\nabla _\sigma }} \right)^*}$ in the second equality of Eq.(\ref{a38}) comes from the expanding of $\phi(\mathbf{X},t)$ or $\phi(\mathbf{x}-\bm{\rho}_0,t)$. Here, to get the approximation,  $O(\left\| {{\bm{\rho} _0}\cdot{\nabla _\sigma }} \right\|) = O(1)$ is used.
The density is obtained by integrating $f(\mathbf{z},t)$ out of the velocity space
\begin{eqnarray}\label{a39}
{n_s}\left( {{\bf{x}},t} \right) &=& {n_{s0}}({\bf{x}}) - \frac{{{q_s}}}{{{T_s}}}\left[ {{n_{s0}}\phi \left( {{\bf{x}},t} \right) - \left\langle {\phi \left( {{\bf{x}} - {\bm{\rho} _0}\left( {\bf{z}} \right),t} \right)} \right\rangle } \right] \nonumber \\
 && + \int {{F_{s1}}\left( {{\bf{x}} - {\bm{\rho} _0}\left( {\bf{z}} \right),{\mu _1},{u_1},{\theta _1},t} \right)\frac{{B\left( {\bf{x}} \right)}}{{{m_s}}}d{u_1}d{\mu _1}d{\theta _1}} ,
\end{eqnarray}
with the definition
\begin{equation}\label{a42}
\left\langle {\phi \left( {{\bf{x}}- {\bm{\rho} _0}\left( {\bf{z}} \right),t} \right)} \right\rangle  \equiv \int {\exp \left( { - {\bm{\rho} _0}\left( {\bf{z}} \right)\cdot{\nabla _\sigma }} \right)\phi \left( {{\bf{x}},t} \right)} {F_{s0}}\left( {\bf{z}} \right)\frac{B(\mathbf{x})}{{{m_s}}}d{u_1}d{\mu _1}d\theta_1.
\end{equation}
Here, $\frac{B(\mathbf{x})}{m_s}$ is the Jacobian determinant between $\bf{v}$ in the rectangular coordinates and $\mu_1,u_1,\theta_1$, while the Jacobian determinant between $\mathbf{z}$ and $\mathbf{Z}$ is approximated to equal one.
The Poisson equation becomes
\begin{equation}\label{a40}
- {\nabla ^2}\phi \left( {{\bf{x}},t} \right) = \frac{1}{\epsilon}\sum\limits_s {{q_s}\left[ \begin{array}{l}
 - \frac{{{q_s}}}{{{T_s}}}\left( {{n_{s0}}\phi \left( {{\bf{x}},t} \right) - \left\langle {\phi \left( {{\bf{x}} - {\bm{\rho} _0}\left( {\bf{z}} \right),t} \right)} \right\rangle } \right)+n_{s1}
\end{array} \right]},
\end{equation}
with the density ${n}_{s1}$ defined as
\begin{equation}\label{a44}
{n_{s1}} \equiv \int {{F_{s1}}\left( {{\bf{x}} - {\bm{\rho} _0}\left( {\bf{z}} \right),{\mu _1},{u_1},{\theta _1},t} \right)\frac{{B\left( {\bf{x}} \right)}}{{{m_s}}}} d{u_1}d{\mu _1}d{\theta _1}.
\end{equation}

We consider a simple plasma which only includes proton and electron. The distribution of electron uses the adiabatic one. Then, the quasi-neutral equation of this plasma is
\begin{equation}\label{a43}
- \frac{e}{{{T_i}}}\left[ {{n_0}\left( {\bf{x}} \right)\phi \left( {{\bf{x}},t} \right) - \left\langle {\phi \left( {{\bf{x}} - {\bm{\rho} _0}\left( {\bf{z}} \right),t} \right)} \right\rangle } \right] + {n_{i1}} - \frac{{{e}{n_0}\left( {\bf{x}} \right)}}{{{T_e}}}\phi \left( {{\bf{x}},t} \right) = 0
\end{equation}

So far, we completed the derivation of the new electrostatic gyrokinetic model, which comprises Eqs.(\ref{a32},\ref{a33},\ref{a34},\ref{a40}).

\section{Summary}\label{sec6}

This paper presented two different electrostatic gyrokinetic models derived through two different methods. A term in the CEGM is rectified in our derivation. Compared with CEGM, the FLR terms are completely removed from the orbit equations in the new electrostatic gyrokinetic model. The reason for the difference is as follows. To derive the coordinate transform of CEGM, terms  contributing the secular terms to the gauge function $S_1$ on the right hand side of Eq.(\ref{g15}) need to be removed, since these secular terms contribute unlimited values to the generators through Eqs. (\ref{g10},\ref{g12},\ref{g20},\ref{g13}), which causes the coordinate transform unacceptable. This operation leaves behind FLR terms to the orbit equations. However, such a situation doesn't exist in deriving the coordinate transform by the new method. The removing of those FLR terms in the orbit equations makes the new model more concise for the numerical application, compared with the conventional one.

\section{Acknowledgement}

This work is partially supported by Grants-in-Aid from JSPS (No.25287153 and 26400531) and by the CSC Scholarship. The author sincerely appreciates the hospitality given by Prof. Yasuaki Kishimoto's laboratory.

\appendix

\section{The non-zero components of the Lie derivatives on $\Gamma_0$ in Eq.(\ref{g8})}\label{app1}

The formula of the Lie derivative of the generators on the differential 1-form is given as
\begin{equation}\label{c13}
{L_{\bf{g}}}\gamma  = \left( {{g^a}{\omega _{ab}} + {\partial _b}\left( {{g^a}{\gamma _a}} \right)} \right)d{z^b}.
\end{equation}
The generator vector $\mathbf{g}$ is defined as ${\bf{g}} \equiv ({{\bf{g}}^{\bf{x}}},{g^\mu },{g^U},{g^\theta })$ with ${{\bf{g}}^{\bf{x}}} = ({g^1},{g^2},{g^3})$ for the spatial space. $g^\mu$,$g^U$ and $g^\theta$ are for the dimensions of $\mu,U,\theta$, respectively. The nonzero components of the Lie derivative on $\Gamma_0$ in Eq.(\ref{g8}) are given below.
\begin{subequations}
\begin{eqnarray}
g_{}^i{\omega _{0ij}}d{X^j}
&=& \left( {{\bf{B}} + \varepsilon U\nabla  \times b} \right) \times {\bf{g}}^\mathbf{x} \cdot d{\bf{X}}, \\
g_{}^U{\omega _{0Ui}}d{X^i}
&=& \varepsilon g_{}^U{\bf{b}}\cdot d{\bf{X}}, \\
g_{}^i{\omega _{0iU}}dU
&=&  - \varepsilon \left( {{\bf{g}}^\mathbf{x}\cdot{\bf{b}}} \right)dU, \\
g_{}^\mu {\omega _{0\mu \theta }}d\theta
&=& {\varepsilon ^2}g_{}^\mu d\theta,  \\
g_{}^\theta {\omega _{0\theta \mu }}d\mu
&=&  - {\varepsilon ^2}g_{}^\theta d\mu,  \\
g_{}^j{\omega _{0jt}}dt
&=&  - \varepsilon \mu {\bf{g}}^\mathbf{x}\cdot\nabla B\left( {\bf{X}} \right)dt,  \\
g_{}^\mu {\omega _{0\mu t}}dt
&=&  - \varepsilon B\left( {\bf{X}} \right)g_{}^\mu dt,  \\
g_{}^U{\omega _{0Ut}}dt
&=&  - \varepsilon Ug_{}^Udt.  \\
\end{eqnarray}
\end{subequations}




%

%

%

%
%

%


\end{document}